\documentclass[aps,pre,preprint,amsmath,amssymb]{revtex4}

\usepackage{graphicx}
\usepackage{dcolumn}
\usepackage{bm}

\usepackage{amsmath}
\usepackage{placeins}

\newcommand{\mbi}[1]{\mbox{\boldmath $#1$}}
\newcommand{\sgn}{\mbox{sgn}}

\begin{document}
%%%%%%%%%%%%%%%%%%%%%%%%%%%%%%%%%%%%%%%%%%%%%%%%%%%%%%%%%%%%%%%%%%%%%%%%%%%%%%%%%%%%%%%%%%%%%
\preprint{}
\title{Statistical mechanics of error correcting code using monotonic and non-monotonic tree-like multilayer perceptrons}
\author{Florent Cousseau}
\affiliation{
Graduate School of Frontier Sciences, University of Tokyo, Chiba 277-5861, Japan}
\email{florent@mns.k.u-tokyo.ac.jp}
\author{Kazushi Mimura}
\affiliation{
Faculty of Information Sciences, Hiroshima City University, Hiroshima 731-3194, Japan}
\email{mimura@hiroshima-cu.ac.jp}
\author{Masato Okada}
\affiliation{
Graduate School of Frontier Sciences, University of Tokyo, Chiba 277-5861, Japan \\
Brain Science Institute, RIKEN, Saitama 351-0198, Japan
}
%%%%%%%%%%%%%%%%%%%%%%%
\date{\today}% It is always \today, today,

\begin{abstract}
\label{abst}
An error correcting code using a tree-like multilayer perceptron
is proposed. An original message $\mbi{s}^0$ is encoded into a codeword $\mbi{y}_0$
using a tree-like committee machine (committee tree) or a tree-like parity machine 
(parity tree). Based on these architectures, several schemes featuring monotonic or non-monotonic units are introduced.
The codeword $\mbi{y}_0$ is then transmitted via a Binary Asymmetric Channel (BAC) 
where it is corrupted by noise. 
The analytical performance of these schemes is investigated using the replica method of statistical mechanics. Under some specific conditions, some of the proposed schemes are shown to saturate the Shannon bound at the infinite codeword length limit. The influence of the monotonicity of the units on the performance is also discussed.
\end{abstract}
%%%%%%%%%%%%%%%%%%%%%%%%%%%%%%%%%%%%%%%%%%%%%%%%%%%%%%%%%%%%%%%%%%%%%%%%%%%%%%%%%%%%%%%%%%%%%

\maketitle
%%%%%%%%%%%%%%%%%%%%%%%%%%%%%%%%%%%%%%%%%%%%%%%%%%%%%%%%%%%%%%%%%%%%%%%%%%%%%%%%%%%%%%%%%%%%%

%%%%%%%%%%%%%%%%%%%%%%%%%%%%%%%%%%%%%%%%%%%%%%%%%%%%%%%%%%%%%%%%%%%%%%%%%%%%%%%%%%%%%%%%%%%%%
\section{introduction}

Reliability in communication has always been a major concern when dealing with digital data. Especially in today's information-dependent society, it is vital to design efficient ways of preventing data corruption when transmitting information. Error correcting codes have been developed for this purpose since the birth of the information theory field following the work of Shannon \cite{Shannon1948}. 

In 1989, Sourlas derived a set of error correcting codes, the so called Sourlas codes, which theoretically saturate the Shannon bound \cite{Sourlas1989}. Although these codes turned out to be impractical, the main point of interest of this paper was the parallel made between physical spin glass systems and information theory.

Following this paper, the tools of statistical mechanics have been successfully applied to a wide range of problems of information theory in recent years. In the field of error correcting codes itself \cite{Kabashima2000,Nishimori1999,Montanari2000},
as well as in spreading codes \cite{Tanaka2001,Tanaka2005},
and compression codes \cite{Murayama2003,Murayama2004,Hosaka2002,Hosaka2005,Mimura2006,Cousseau2008}, statistical mechanical techniques have shown great potential.

The present paper uses similar techniques to investigate an error correcting code scheme where the codeword is encoded using tree-like multilayer perceptron neural networks. It is known that there exists a natural duality between lossy compression codes and error correcting codes. Indeed, a lossy compression code can be regarded as a standard error correcting code, but one where the codeword is generated using the original decoder of the error correcting code scheme and where the decompressed message is obtained using the original encoder of the scheme (Cf. \cite{MacKay2003} for details).

Recently, a lossy compression scheme based on a simple perceptron decoder was investigated by Hosaka et al. \cite{Hosaka2002}. In their paper, they used statistical mechanical techniques to investigate the theoretical performance of their scheme at the infinite codeword length limit. The perceptron they defined in their model uses a special hat-shaped non-monotonic transfer function. This rather uncommon feature enables the scheme to deal with biased messages and it is known that this type of function maximizes the storage capacity of the simple perceptron \cite{Monasson1994,Bex1995}. They found that their scheme can theoretically yield Shannon optimal performance. Subsequently, Shinzato et al. \cite{Shinzato2006} investigated the same model but in the framework of error correcting code. They found that their model can theoretically yield Shannon optimal performance.

Based on these studies, Mimura et al. \cite{Mimura2006} proposed a tree-like multilayer perceptron network for lossy compression purposes, but use only the standard sign function as the transfer function of their model. They showed that the parity tree model can theoretically yield Shannon optimal performance, but only when considering unbiased messages. In contrast, they showed that the committee tree model cannot yield optimal performance, even for unbiased messages. However, the advantage of using a multilayer structure is improved replica symmetric solution stability, and an increased number of codewords sharing the same distortion properties \cite{Almeida1978}.
In a recent study, Cousseau et al. \cite{Cousseau2008} investigated the same tree-like multilayer perceptron model but used the hat-shaped non-monotonic transfer function introduced by Hosaka et al. \cite{Hosaka2002}, thus combining both advantages of \cite{Hosaka2002,Mimura2006}. By doing so, they were able to show that both parity tree and committee tree structures can then theoretically yield Shannon optimal performance even for biased messages under some specific conditions.

The purpose of the present paper is to discuss the performance of the same tree-like perceptron models but in the error correcting code framework, thus completing the topic of perceptron type network applications in coding theory. In this paper, we make use of the Binary Asymmetric Channel (BAC). Indeed, the use of the non-monotonic hat-shaped transfer function introduced by Hosaka et al. \cite{Hosaka2002} enables us to control the bias of the of the codeword sequence, and enables the relevant schemes to deal with such an asymmetric channel (the BAC was also used by Shinzato et al. \cite{Shinzato2006}). On the other hand, we expect the schemes which use the standard monotonic sign function to be able to deal only with the BSC channel, which corresponds to a particular case of the BAC. The majority of popular error correcting codes like turbo codes \cite{Berrou1993} and low density parity check codes (LDPC) \cite{Gallager1962,MacKay1995}, which provide near Shannon performance in practical time frames, have been widely studied but this was generally restricted to symmetric channels. On the other hand, apart from a few studies \cite{Wang2005,Neri2008}, little is known when dealing with asymmetric channels.

Multilayer perceptrons have been widely studied over the years by the machine learning community and a wide range of problems have been considered (storage capacity, learning rules, etc). These works revealed non-trivial behaviors of even simple models like the simple perceptron network for example. Many of these previous results are summarized in reference \cite{Engel2001}. The present analysis gives us an opportunity to discuss the difficulty of decoding for densely connected systems (or dense systems as opposed to sparsely connected systems like LDPC codes for example) using a systematic manner in the context of multilayer networks. There has been relatively little discussion of dense systems, mainly because of the computational cost which is obviously higher than for sparse systems. However, because of their their rich randomness, dense systems can possibly be regarded as pseudo-random codes like the dense limit of LDPC codes.

In this paper we mainly focus on the necessary conditions to get Shannon optimal performance. To discuss practical decoders, it is first necessary to investigate the optimality of our schemes. This includes discussion of the optimal parameters for the transfer function since we need to know these parameters to discuss the optimal decoder. In other words, we need a theoretical analysis of the performance before we can study the decoding problem. 

The paper is organized as follows. Section \ref{section2} introduces the framework of
error correcting codes. Section \ref{section3} describes our model. Section \ref{section4} deals with the BAC capacity.
Section \ref{section5} presents the mathematical tools used to evaluate the performance of the present scheme. Section \ref{section6} states the results and elucidates the location of the phase transition, which characterizes the best achievable performance of the model. Section \ref{section7} is devoted to the conclusion and discussion.

\section{Error correcting codes} \label{section2}

In a general scheme, an original message $\mbi{s}^0$ of size $N$ is encoded into a codeword $\mbi{y}_0$ of size $M$ by some encoding device. The aim of this stage is to add redundancy to the original data. Therefore, we necessarily have $M>N$. Based on this redundancy, a proper decoder device should be able to recover the original data even if it were corrupted by noise in the transmission channel. The quantity $R=N/M$ is called the code rate and evaluates the trade-off between redundancy and codeword size. The codeword $\mbi{y}_0$ is then fed into a channel where the bits are subject to noise. The received noisy message $\mbi{y}$ (which is also $M$ dimensional) is then decoded using its redundancy to infer the original $N$ dimensional message $\mbi{s}^0$. In other words, in a Bayesian framework, one tries to maximize the following posterior probability,
\begin{eqnarray}
P(\mbi{s}|\mbi{y}) \propto P(\mbi{y}|\mbi{s}) P(\mbi{s}).
\end{eqnarray}

As data transmission is costly, generally one wants to be able to ensure error-free transmission while transmitting the fewest possible bits. In other words, one wants to ensure error-free transmission while keeping the code rate as large as possible. For this purpose, the well known Shannon bound \cite{Shannon1948} gives a way to compute the best achievable code rate which allows error-free recovery. However, while this gives us the value of such an optimal code rate, it does not give any clue as to how to construct such an optimal code. Therefore, several codes have been proposed over the years in an ongoing quest to find a code which can reach this theoretical bound.

\section{Error correcting codes using monotonic and non-monotonic multilayer perceptrons} \label{section3}

In this paper, since we make use of techniques derived from statistical mechanics, we will use Ising variables rather than Boolean ones. The Boolean $0$ is mapped onto $1$ in the Ising framework while the Boolean $1$ is mapped to $-1$. This mapping can be used without any loss of generality.

\begin{figure}[ht!]
  \vspace{0mm}%<--space
  \begin{center}
  \includegraphics[width=0.3\linewidth,keepaspectratio]{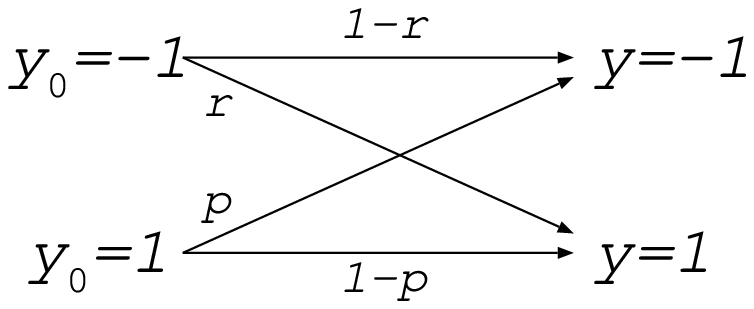}
  \end{center}
  \caption{The Binary Asymmetric Channel (BAC)}
  \label{fig:BAC}
  \vspace{0mm}%<--space
\end{figure}

We assume that the original message $\mbi{s}_0$ is generated from the uniform distribution and that all the bits are independently generated so that we have 
\begin{eqnarray}
P(\mbi{s}^0)= \frac1 {2^N}.
\end{eqnarray}
The channel considered in this study is the Binary Asymmetric Channel (BAC) where each bit is flipped independently of the others with asymmetric probabilities. If the original bit fed into the channel is $1$, then it is flipped with probability $p$. Conversely, if the original bit is $-1$, it is flipped with probability $r$. Figure \ref{fig:BAC} shows the BAC properties in details.
The well known Binary Symmetric Channel (BSC) corresponds to the particular case $r=p$.

\begin{figure}[hb!]
  \vspace{0mm}%<--space
  \begin{center}
  \includegraphics[width=0.5\linewidth,keepaspectratio]{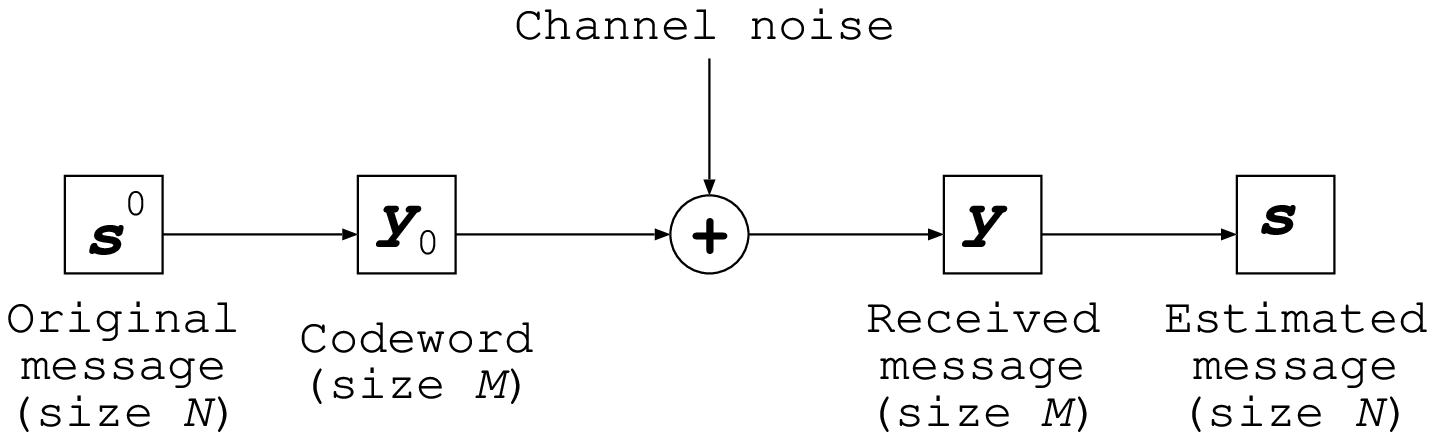}
  \end{center}
  \caption{Layout of the scheme}
  \label{fig:scheme}
  \vspace{0mm}%<--space
\end{figure}

When the corrupted message $\mbi{y}$ is received at the output of the channel, the goal is then to recover $\mbi{s}^0$ using $\mbi{y}$. The state of the estimated message is denoted by the vector $\mbi{s}$. The general outline of the scheme is shown in Figure \ref{fig:scheme}.
From Figure \ref{fig:BAC} we can easily derive the following conditional probability,
\begin{eqnarray}
P(y^{\mu}|y^{\mu}_0) = \frac 1 2 + \frac {y^{\mu}} 2 [(1-r-p) y^{\mu}_0 + (r-p)],
\end{eqnarray}
where we make use of the notations $\mbi{y}_0 = (y_0^1,\ldots,y_0^{\mu},\ldots,y_0^M)$, $\mbi{y} = (y^1,\ldots,y^{\mu},\ldots,y^M)$. Since we assume that the bits are flipped independently, we deduce
\begin{eqnarray}
P(\mbi{y}|\mbi{y}_0) = \prod_{\mu=1}^M \left\{ \frac 1 2 + \frac {y^{\mu}} 2 [(1-r-p) y^{\mu}_0 + (r-p)] \right\}.
\end{eqnarray}

To encode the original message $\mbi{s}^0$ into a codeword $\mbi{y}_0$, we use three non-monotonic tree-like parity machine or committee machine neural networks ((I), (II) and (III)). In the same way, we also investigate the standard monotonic parity tree and committee tree neural networks ((IV) and (V)).

(I) Multilayer parity tree with non-monotonic hidden units (PTH).
\begin{equation}
y^{\mu}_0 (\mbi{s}^0) \equiv
\prod_{l=1}^K f_k \left( \sqrt{\frac {K} {N}} \, \mbi{s}^0_l \cdot \mbi{x}_l^{\mu} \right) .
\label{parity}
\end{equation}

(II) Multilayer committee tree with non-monotonic hidden units (CTH).
\begin{equation}
y^{\mu}_0 (\mbi{s}^0) \equiv
\sgn \left( \sum_{l=1}^K f_k \left[ \sqrt{\frac {K} {N}} \, \mbi{s}^0_l \cdot \mbi{x}_l^{\mu} \right] \right) .
\label{committee1}
\end{equation}
Note that in this case, if the number of hidden units $K$ is even, it is possible to get $0$ as the argument of the sign
function. We avoid this uncertainty by considering only an odd
number of hidden units for the committee tree with non-monotonic
hidden units in the sequel.

(III) Multilayer committee tree with a non-monotonic output unit (CTO).
\begin{equation}
y^{\mu}_0 (\mbi{s}^0) \equiv
f_k \left( \sqrt {\frac 1 K}\sum_{l=1}^K
\sgn \left[ \sqrt{\frac {K} {N}} \, \mbi{s}^0_l \cdot \mbi{x}_l^{\mu}
\right] \right) .
\label{committee2}
\end{equation}

(IV) Multilayer parity tree (PT).
\begin{equation}
y^{\mu}_0 (\mbi{s}^0) \equiv
\prod_{l=1}^K \sgn \left( \sqrt{\frac {K} {N}} \, \mbi{s}^0_l \cdot \mbi{x}_l^{\mu} \right) .
\label{MONOparity}
\end{equation}

(V) Multilayer committee tree (CT).
\begin{equation}
y^{\mu}_0 (\mbi{s}^0) \equiv
\sgn \left( \sqrt {\frac 1 K} \sum_{l=1}^K \sgn \left[ \sqrt{\frac {K} {N}} \, \mbi{s}^0_l \cdot \mbi{x}_l^{\mu} \right] \right) .
\label{MONOcommittee}
\end{equation}
In this case also, if the number of hidden units $K$ is even,
it is a possible to get $0$ as the argument of the sign
function. We again avoid this uncertainty by considering only an odd
number of hidden units for the committee tree in the sequel.

\begin{figure}[ht!]
  \vspace{0mm}%<--space
  \begin{center}
  \includegraphics[width=0.6\linewidth,keepaspectratio]{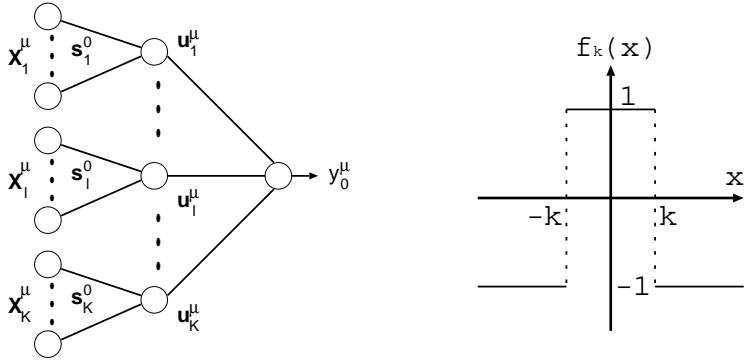}
  \end{center}
  \caption{Left: General architecture of the treelike multilayer perceptrons with $N$ input units and $K$ hidden units. Right: The non-monotonic function $f_k$.}
  \label{fig:net}
  \vspace{0mm}%<--space
\end{figure}

The original message $\mbi{s}^0$ is split into $N / K$-dimensional $K$ disjoint vectors so that $\mbi{s}^0$ can be written $\mbi{s}^0=(\mbi{s}^0_1, \ldots , \mbi{s}^0_K)$. In schemes (I), (II), and (III), $f_k$ is a non-monotonic function of a real parameter $k$ of the
form
\begin{equation}
f_k (x) =
\begin{cases}
1 \quad \text{if} \quad |x| \leq k
\\
-1 \quad \text{if} \quad |x| > k ,
\end{cases}
\end{equation}
and the vectors $\mbi{x}^{\mu}_l$ are fixed $N/K$-dimensional independent vectors
uniformly distributed on $\{-1,1\}$. The use of random input vectors is known to maximize the storage capacity of perceptron networks, making such a scheme promising for error correcting tasks. The $\sgn$ function denotes the sign function taking $1$
for $x \geq 0$ and $-1$ for $x <0$. Each of these architectures applies a
different non-linear transformation to the original data $\mbi{s}^0$.
The general architecture of these perceptron-based encoders and the non-monotonic function $f_k$ are displayed in Figure \ref{fig:net}.
Note that we can also consider an encoder based on a committee-tree where both the hidden-units
and the output unit are non-monotonic. However, this introduces an extra parameter (we will
have one threshold parameter for the hidden-units and one for the output unit) to tune
and the performance should not change drastically. For simplicity, we restrict our study
to the above three cases.

To keep the notation as general as possible, as long as explicit use of the encoder is not necessary in computations, we will denote the transformation performed on vector $\mbi{s}$ by the respective encoders using the following notation:
\begin{equation}
\mathcal{F}_k \left( \left\{ \sqrt{\frac K N} \mbi{s}_l \cdot \mbi{x}_l^{\mu} \right\} \right).
\end{equation}
$\mathcal{F}_k$ takes a different expression for the five different types of network and $k$ denotes the fact that all the encoders depend on a real threshold parameter $k$ (except for schemes (IV) and (V), where this function does not depend on $k$. However for consistency, we will keep this notation for these schemes). Furthermore, note that $\mathcal{F}_k$ contains all the terms depending on index $l$ (i.e.: $\mathcal{F}_k ( \{ u_l \} )$ contains all the terms $u_1, \ldots , u_l, \ldots , u_K$).

\section{Binary Asymmetric Channel (BAC) capacity} \label{section4}

In this section, we compute the capacity of the BAC. According to Shannon's channel coding theorem, the optimal code rate is given by the capacity of the channel. Any code rate bigger than the channel capacity will inevitably lead to information loss. The definition of the channel capacity $C$ is 
\begin{eqnarray}
C = \max_{{\rm{input \ probability}}} \left\{ I(X,Y) \right\},
\end{eqnarray}
where $I$ denotes mutual information, $X$ denotes the channel input distribution, and $Y$ denotes the channel output distribution. Computation of the capacity of such a binary channel requires only simple algebra and calculations are straightforward, giving
\begin{eqnarray}
C_{BAC} = H_2 (\gamma_C) - \frac {1+\Omega_C} 2 H_2 (p) - \frac {1-\Omega_C} 2 H_2 (r), \label{capacity}
\end{eqnarray}
where
\begin{eqnarray}
H_2 (x) & = & - x \log_2 (x) - (1-x) \log_2 (1-x), \\
\gamma_C & = & \frac {1} {1+\Delta_C} = \frac 1 2 \left[ (1-p) (1+\Omega_C) + r (1-\Omega_C) \right], \\
\Delta_C & = & \left[ \frac {r^r (1-r)^{1-r}} {p^p (1-p)^{1-p}} \right]^{1/1-r-p}, \\
\Omega_C & = & \frac {2 \gamma_C - 1 -r + p} {1 -r -p}. \label{omegC}
\end{eqnarray}
In the special case $r=p$, the capacity simplifies to
\begin{eqnarray}
C_{BSC} = 1 - H_2 (p),
\end{eqnarray}
which corresponds to the capacity of the BSC.

\section{Analytical Evaluation} \label{section5}

As stated in section II, our goal is to maximize the posterior $P(\mbi{s}|\mbi{y})$. 
Let us define the following Hamiltonian:
\begin{eqnarray}
\mathcal{H} (\mbi{y},\mbi{s}) = - \ln [ P(\mbi{s}|\mbi{y}) P(\mbi{s}) ] 
= - \ln P(\mbi{y},\mbi{s}).
\end{eqnarray}
The ground state of the above Hamiltonian trivially corresponds to the {\it maximum a posteriori} (MAP) estimator of the posterior $P(\mbi{s}|\mbi{y})$. Then, let us compute the joint probability of $\mbi{y}$ and $\mbi{s}$. We have
\begin{eqnarray}
P(\mbi{y},\mbi{s}) = P(\mbi{y}|\mbi{s}) P(\mbi{s}).
\end{eqnarray}
Since the relation between an arbitrary message $\mbi{s}$ and the codeword fed into the channel is deterministic, for any $\mbi{s}$, we can write
\begin{eqnarray}
P(\mbi{y}|\mbi{s}) & = & P \left( \mbi{y} \Big| \mathcal{F}_k \left( \left\{ \sqrt{\frac K N} \mbi{s}_l \cdot \mbi{x}_l^{\mu} \right\} \right) \right), \nonumber
\\
& = & \prod_{\mu=1}^M \left\{ \frac 1 2 + \frac {y^{\mu}} 2 [(1-r-p) \mathcal{F}_k \left( \left\{ \sqrt{\frac K N} \mbi{s}_l \cdot \mbi{x}_l^{\mu} \right\} \right) + (r-p)] \right\}.
\end{eqnarray}
We finally get the explicit expression of the Hamiltonian,
\begin{eqnarray}
\mathcal{H} (\mbi{y},\mbi{s}) & = & - \ln P(\mbi{y},\mbi{s}) \nonumber
\\
& = & - \ln \left[ \frac 1 {2^N} \prod_{\mu=1}^M \left\{ \frac 1 2 + \frac {y^{\mu}} 2 [(1-r-p) \mathcal{F}_k \left( \left\{ \sqrt{\frac K N} \mbi{s}_l \cdot \mbi{x}_l^{\mu} \right\} \right) + (r-p)] \right\} \right] .
\end{eqnarray}
Using this Hamiltonian, we can define the following partition function
\begin{equation}
Z(\beta,\mbi{y},\mbi{x}) = \sum_{\mbi{s}}
\exp \left[ -\beta \mathcal{H} (\mbi{y},\mbi{s}) \right] ,
\end{equation}
where the sum over $\mbi{s}$ represents the sum over all possible states for vector
$\mbi{s}$, and $\beta$ is the inverse temperature parameter. Such a partition function
can be identified with the partition function of a spin glass system with dynamical
variables $\mbi{s}$ and quenched variables $\mbi{x}$. The average of this partition
function over $\mbi{y}$ and $\mbi{x}$ naturally contains all the interesting typical
properties of the scheme, such as the free energy. However, it is hard to evaluate this average and we need some techniques to investigate it. In this paper, we use the so-called
\textit{Replica Method} to calculate the average of the partition function.
Once the free energy is obtained, one can compute the critical code rate at which a phase transition occurs between the ferromagnetic phase (error recovery possible) and the paramagnetic phase (decoding impossible). This gives us the best code rate the scheme can achieve. A code rate exceeding this critical value will make decoding impossible.
The calculations to obtain the average of the partition function
$\left< Z(\beta,\mbi{y},\mbi{x}) \right>_{\mbi{y},\mbi{x}}$ are detailed in Appendix A.

After long calculations, the replica symmetric (RS) free energy is obtained,
\begin{eqnarray}
- f_{RS} (q,\hat{q},m,\hat{m}) & = & \underset{q,\hat{q},m,\hat{m}}{{\rm{extr}}} \left\{
\sum_{y=\pm1} \int_{-\infty}^{\infty} \left[ \prod_{l=1}^{K} D R_l \right] \int_{-\infty}^{\infty} \left[ \prod_{l=1}^{K} D t_l \right] \times \ln \left[ I(y,R_l,t_l,m,q) \right] \right. \nonumber
\\
&& \times \left( \frac 1 2 + \frac y 2 \left[ (1-r-p) \mathcal{F}_k \left( \left\{ R_l \right\} \right) + (r-p) \right] \right) \nonumber
\\
&& + R \int_{-\infty}^{\infty} DU \ln \left( 2 \cosh \left[ \sqrt{\hat{q}} U + \hat{m} \right] \right) - R \ln 2 \nonumber
\\
&& - R m \hat{m} -R \frac { \hat{q} (1-q)} 2  \Bigg\} , \label{freeEnergy}
\end{eqnarray}
where
\begin{eqnarray}
I(y,R_l,t_l,m,q) & = & \int_{-\infty}^{\infty} \left[ \prod_{l=1}^{K} D z_l \right] 
\times \Bigg[ \frac 1 2 + \frac y 2 (r-p) 
\nonumber \\ && 
+\frac y 2 (1-r-p)
\mathcal{F}_k \left( \left\{ \sqrt{1-q} z_l + \sqrt{q-m^2} t_l +m R_l \right\} \right) \Bigg],
\end{eqnarray}
\begin{eqnarray}
Dx & = & \frac {e^{- \frac {x^2} 2}} {\sqrt{2 \pi}} dx.
\end{eqnarray}
and where ${\rm{extr}}$ denotes extremization. The sum denotes the sum other all possible states for the variable $y$, that is $\pm 1$. 

Note also that we set $\beta=1$. This choice of finite temperature decoding (in contrast to $\beta \to \infty$ which corresponds to the zero temperature limit) corresponds to the {\it maximizer of posterior marginals} (MPM) estimator, while the zero temperature decoding corresponds to the MAP estimator \cite{Rujan1993,Nishimori2001}. The MPM estimator is known to be optimal for the purpose of decoding \cite{Nishimori1993,Sourlas1994,Nishimori2001}. On top of that, in this paper we suppose that all the channel properties (i.e.: the true values of $(p,r)$) are known to the decoder which implies that the system's state we consider is located on the Nishimori line \cite{Nishimori1993,Sourlas1994}.

To retrieve the free energy one has to extremize (\ref{freeEnergy}) with respect to the order parameters $q,\hat{q},m,\hat{m}$. This is done by solving the following saddle point equations
\begin{eqnarray}
\frac {\partial f_{RS}} {\partial q} & = & 0 \Leftrightarrow \hat{q}=-2 R^{-1}
\sum_{y=\pm1} \int_{-\infty}^{\infty} \left[ \prod_{l=1}^{K} D R_l \right] \int_{-\infty}^{\infty} \left[ \prod_{l=1}^{K} D t_l \right] \times \frac{I^{\prime}_q (y,R_l,t_l,m,q)} {I(y,R_l,t_l,m,q)} \nonumber
\\
&& \makebox[4cm]{} \times \left( \frac 1 2 + \frac y 2 \left[ (1-r-p) \mathcal{F}_k \left( \left\{ R_l \right\} \right) + (r-p) \right] \right), \label{qHat}
\\
\frac {\partial f_{RS}} {\partial m} & = & 0 \Leftrightarrow \hat{m}= R^{-1}
\sum_{y=\pm1} \int_{-\infty}^{\infty} \left[ \prod_{l=1}^{K} D R_l \right] \int_{-\infty}^{\infty} \left[ \prod_{l=1}^{K} D t_l \right] \times \frac{I^{\prime}_m (y,R_l,t_l,m,q)} {I(y,R_l,t_l,m,q)} \nonumber
\\
&& \makebox[4cm]{} \times \left( \frac 1 2 + \frac y 2 \left[ (1-r-p) \mathcal{F}_k \left( \left\{ R_l \right\} \right) + (r-p) \right] \right), \label{mHat}
\\
\frac {\partial f_{RS}} {\partial \hat{q}} & = & 0 \Leftrightarrow 
q = \int_{-\infty}^{\infty} D U \tanh^2 (\sqrt{\hat{q}} U + \hat{m}), \label{q}
\\
\frac {\partial f_{RS}} {\partial \hat{m}} & = & 0 \Leftrightarrow \label{m}
m = \int_{-\infty}^{\infty} D U \tanh (\sqrt{\hat{q}} U + \hat{m}),
\end{eqnarray}
where
\begin{eqnarray}
I^{\prime}_q (y,R_l,t_l,m,q) & = & \frac {\partial I (y,R_l,t_l,m,q)} {\partial q},
\\
I^{\prime}_m (y,R_l,t_l,m,q) & = & \frac {\partial I (y,R_l,t_l,m,q)} {\partial m}.
\end{eqnarray}
An error correcting code scheme typically admits two solutions: one where $m=q=1$, called the ferromagnetic solution, and one where $m=q=0$, called the paramagnetic solution. As the names indicate, these solutions come from the physical ferromagnet state and correspond to the case where the spins are all ordered ($m=q=1$) or to the case where the spins take completely random states ($m=q=0$). As we can deduce from equations (\ref{mOrder}) and (\ref{RS}), the ferromagnetic solution corresponds to decoding success since $m=1$ implies perfect overlap. Conversely, the paramagnetic phase implies failure in the decoding process (overlap $m$ is $0$).

\subsection{Replica symmetric solution using a parity tree with non-monotonic hidden units}

Using a parity tree with non-monotonic hidden units (\ref{parity}), the encoder function becomes
\begin{eqnarray}
\mathcal{F}_k ( \{ u_l \} ) = \prod^K_{l=1} f_k (u_l).
\end{eqnarray}
Using this encoder function and substituting $m=q=0$ in the saddle point equations, one can find a consistent solution where $q=m=\hat{q}=\hat{m}=0$. This corresponds to the paramagnetic solution, where decoding of the received message fails. Using these conditions in (\ref{freeEnergy}), one can retrieve the free energy of the paramagnetic phase,
\begin{eqnarray}
- f_{para} & = & - H_2 \left(\frac 1 2 \left[ (1-p) (1+\Omega_{PTH}) + r(1-\Omega_{PTH}) \right] \right) \times \ln 2 , \label{freePara}
\end{eqnarray}
where
\begin{eqnarray}
\Omega_{PTH} = \prod_{l=1}^K \int_{-\infty}^{+\infty} D z_l f_k(z_l). \label{OmegaPT}
\end{eqnarray}
In the same way, substituting $m=q=1$ in the saddle point equations, one can find a consistent solution. However, the ferromagnetic solution cannot be computed analytically. So we proceed numerically by simply checking the integrand of equations (\ref{qHat}) and (\ref{mHat}). We did that extensively for values of $K=1$, $K=2$, and $K=3$. In each case we found that the integrand diverges so that when $(q,m)\to(1,1)$, we have both $\hat{q}\to\infty$ and $\hat{m}\to\infty$. Substituting $\hat{q}\to\infty$ and $\hat{m}\to\infty$ into (\ref{q}) and (\ref{m}) clearly yields $q=m=1$. So $q=m=1$, $\hat{q}\to\infty$ and $\hat{m}\to\infty$ is a consistent solution of the saddle point equations which corresponds to the ferromagnetic solution, where decoding of the received message succeeds. We also checked higher values of $K$ (up to $K=5$) and did not find any other consistent solution. We conjecture that this result holds for any finite value of $K$. Finally, substituting $m=q=1$, $\hat{m} \to \infty$ and $\hat{q} \to \infty$ into (\ref{freeEnergy}), one can get the free energy of the ferromagnetic phase,
\begin{eqnarray}
- f_{ferro} & = & - \frac {\ln 2} {2} \left[ (1+\Omega_{PTH}) H_2 (p) + (1-\Omega_{PTH}) H_2 (r) \right] - R \ln 2. \label{freeFerro}
\end{eqnarray}
Note that when $K=1$, the present scheme corresponds to the case of Shinzato et al. \cite{Shinzato2006}. The result we obtained when $K=1$ is indeed equivalent to what they found.

\subsection{Replica symmetric solution using a committee tree with non-monotonic hidden units}

When a committee tree with non-monotonic hidden units (\ref{committee1}) is used, the encoder function becomes
\begin{eqnarray}
\mathcal{F}_k ( \{ u_l \} ) = \sgn \left[ \sum^K_{l=1} f_k (u_l) \right].
\end{eqnarray}
Using this encoder function and substituting $m=q=0$ in the saddle point equations, one can find a consistent solution where $q=m=\hat{q}=\hat{m}=0$. This corresponds to the paramagnetic solution, where decoding of the received message fails. Using these conditions in (\ref{freeEnergy}), one can retrieve the free energy of the paramagnetic phase,
\begin{eqnarray}
- f_{para} & = & - H_2 \left(\frac 1 2 \left[ (1-p) (1+\Omega_{CTH}) + r(1-\Omega_{CTH}) \right] \right) \times \ln 2 ,
\end{eqnarray}
where
\begin{eqnarray}
\Omega_{CTH} = \int_{-\infty}^{+\infty} \left[ \prod_{l=1}^K D z_l \right] \times \sgn \left[ \sum^K_{l=1} f_k (z_l) \right]. \label{OmegaCTH}
\end{eqnarray}
In the same way, by substituting $m=q=1$ in the saddle point equations one can find a consistent solution. However, the ferromagnetic solution cannot be computed analytically, so we proceed numerically by simply checking the integrand of equations (\ref{qHat}) and (\ref{mHat}). We did that extensively for $K=3$ (we consider only odd values of $K$ for this scheme, and when $K=1$ the present scheme is equivalent to the parity tree case). We found that the integrand diverges so that when $(q,m)\to(1,1)$, we have both $\hat{q}\to\infty$ and $\hat{m}\to\infty$. We also checked higher values of $K$ (up to $K=5$) and did not find any other consistent solution. We conjecture that this result holds for any finite value of $K$. Finally, substituting $m=q=1$, $\hat{m} \to \infty$ and $\hat{q} \to \infty$ into (\ref{freeEnergy}), one can get the free energy of the ferromagnetic phase,
\begin{eqnarray}
- f_{ferro} & = & - \frac {\ln 2} {2} \left[ (1+\Omega_{CTH}) H_2 (p) + (1-\Omega_{CTH}) H_2 (r) \right] - R \ln 2.
\end{eqnarray}

\subsection{Replica symmetric solution using a committee tree with a non-monotonic output unit}

When a committee tree with a non-monotonic output unit (\ref{committee2}) is used, the encoder function becomes
\begin{eqnarray}
\mathcal{F}_k ( \{ u_l \}) = f_k \left[ \sqrt{\frac {1} {K}} \sum^K_{l=1} \sgn (u_l) \right].
\end{eqnarray}
Using this encoder function and substituting $m=q=0$ in the saddle point equations do not imply $\hat{m}=\hat{q}=0$ and a non-trivial solution is found, which makes the free energy too complex to be investigated. This scheme is likely to give non-optimal performance in such a case and will not be considered in what follows.

Note that the limit where $K \to \infty$ was not studied because the saddle point equations take a non-trivial form that is difficult to investigate (in the lossy compression case, this study is still tractable). The techniques to investigate the free energy in the $K \to \infty$ limit described in reference \cite{Engel2001} cannot be easily applied here. However, based on the previous results of Cousseau et al. \cite{Cousseau2008}, it is probable that in the $K \to \infty$ limit, the committee tree with a non-monotonic output unit saturates the Shannon bound in the general BAC case.

\subsection{Replica symmetric solution using a parity tree}

Using a parity tree (\ref{MONOparity}), the encoder function becomes
\begin{eqnarray}
\mathcal{F}_k ( \{ u_l \} ) = \prod^K_{l=1} \sgn (u_l).
\end{eqnarray}
Using this encoder function and substituting $m=q=0$ in the saddle point equations, one can find a consistent solution where $q=m=\hat{q}=\hat{m}=0$ but only when $K>1$. This corresponds to the paramagnetic solution, where decoding of the received message fails. Using these conditions in (\ref{freeEnergy}), one can retrieve the free energy of the paramagnetic phase,
\begin{eqnarray}
- f_{para} & = & - H_2 \left(\frac 1 2 \left[ (1-p) (1+\Omega_{PT}) + r(1-\Omega_{PT}) \right] \right) \times \ln 2 ,
\end{eqnarray}
where
\begin{eqnarray}
\Omega_{PT} = \prod_{l=1}^K \int_{-\infty}^{+\infty} D z_l \times \sgn (z_l). \label{OmegaMONOPT}
\end{eqnarray}
When $K=1$ is considered, $m=q=0$ does not imply $\hat{m}=\hat{q}=0$ and a non-trivial solution is found that makes the free energy too complex to be investigated. The scheme is likely to give non-optimal performance in such a case and will not be considered in what follows.

In the same way, substituting $m=q=1$ in the saddle point equations, one can find a consistent solution, but only when $K>1$. However, the ferromagnetic solution cannot be computed analytically, so we proceed numerically by simply checking the integrand of equations (\ref{qHat}) and (\ref{mHat}). We did that extensively for values of $K=2$ and $K=3$. In each case, we found that the integrand diverges so that when $(q,m)\to(1,1)$ we have both $\hat{q}\to\infty$ and $\hat{m}\to\infty$. We also checked higher values of $K$ (up to $K=5$) and did not find any other consistent solution. We conjecture that this result holds for any finite value of $K>1$. Finally, substituting $m=q=1$, $\hat{m} \to \infty$ and $\hat{q} \to \infty$ into (\ref{freeEnergy}), one can get the free energy of the ferromagnetic phase,
\begin{eqnarray}
- f_{ferro} & = & - \frac {\ln 2} {2} \left[ (1+\Omega_{PT}) H_2 (p) + (1-\Omega_{PT}) H_2 (r) \right] - R \ln 2.
\end{eqnarray}

\subsection{Replica symmetric solution using a committee tree}

Using a committee tree (\ref{MONOcommittee}), the encoder function becomes
\begin{eqnarray}
\mathcal{F}_k ( \{ u_l \}) = \sgn \left[ \sqrt{\frac {1} {K}} \sum^K_{l=1} \sgn (u_l) \right].
\end{eqnarray}
Using this encoder function and substituting $m=q=0$ in the saddle point equations do not imply $\hat{m}=\hat{q}=0$ and a non-trivial solution is found that makes the free energy too complex to be investigated. This scheme is likely to give non-optimal performance in such a case and will not be considered in what follows. As in the lossy compression case \cite{Mimura2006}, the committee tree is unable to yield Shannon optimal performance. 

Note that the limit where $K \to \infty$ was not studied because the saddle point equations take a non-trivial form that is difficult to investigate (in the lossy compression case, this study is still tractable). The techniques to investigate the free energy in the $K \to \infty$ limit described in reference \cite{Engel2001} cannot be easily applied here. However, based on the previous results of Mimura et al. \cite{Mimura2006}, it is probable that in the $K \to \infty$ limit the committee tree still fails to saturate the Shannon bound even in the BSC case.

\section{Phase transition} \label{section6}

For the parity and committee tree with non-monotonic hidden units and for the standard parity tree, we found a paramagnetic and a ferromagnetic solution of the following form:
\begin{eqnarray}
- f_{para} & = & - H_2 \left(\frac 1 2 \left[ (1-p) (1+\Omega) + r(1-\Omega) \right] \right) \times \ln 2 ,
\\
- f_{ferro} & = & - \frac {\ln 2} {2} \left[ (1+\Omega) H_2 (p) + (1-\Omega) H_2 (r) \right] - R \ln 2,
\end{eqnarray}
where $\Omega$ is given by $\Omega_{PTH}$, $\Omega_{CTH}$, or $\Omega_{PT}$ depending on the encoder considered.

It then beconmes possible to calculate the critical value of the code rate $R$ at which a sharp phase transition occurs between the ferromagnetic and the paramagnetic phase. This indicates the boundary between possible decoding (ferromagnetic phase) and impossible decoding (paramagnetic phase). In other words, this enables us to calculate the optimal code rate for each scheme. At the phase transition point, we have
\begin{eqnarray}
f_{para} = f_{ferro}.
\end{eqnarray}
Simple algebra leads to
\begin{eqnarray}
R = H_2  (\gamma) - \frac {1+\Omega} {2} H_2 (p) - \frac {1-\Omega} {2} H_2 (r),
\end{eqnarray}
where
\begin{eqnarray}
\gamma = \frac 1 2 \left[ (1-p) (1+\Omega) + r (1-\Omega) \right]
\end{eqnarray}
and where $\Omega$ is given by the encoder considered ($\Omega_{PTH}$, $\Omega_{CTH}$, or $\Omega_{PT}$). This equation has exactly the same form as the BAC capacity equation (\ref{capacity}) and in fact is equivalent to the BAC capacity if and only if $\Omega = \Omega_C$. Since $\Omega$ depends on the encoder, we will treat each case in the following subsections.

\subsection{Tuning of the parity tree with non-monotonic hidden units}
In the parity tree with non-monotonic hidden units case, we have
\begin{eqnarray}
\Omega \equiv \Omega_{PTH} = \prod_{l=1}^K \int_{-\infty}^{+\infty} D z_l f_k(z_l).
\end{eqnarray}
The parity tree with non-monotonic hidden units is optimal if and only if 
\begin{eqnarray}
\Omega_{PTH} = \Omega_{C} \Leftrightarrow H (k) = \frac 1 4 \left( 1 - \sqrt[K]{\Omega_C} \right), \label{kPT}
\end{eqnarray}
where
\begin{eqnarray}
H (x) = \int_{x}^{+\infty} Dz.
\end{eqnarray}
This gives us a condition on the threshold parameter $k$ of the non-monotonic transfer function $f_k$. If the threshold $k$ is tuned to satisfy (\ref{kPT}), the scheme achieves the Shannon limit. The only remaining issue is whether such an optimal threshold $k$ exists. 

We solved (\ref{kPT}) numerically with parameters $(p,r) \in \{]0,1[\}^2$ and always found an optimal threshold parameter $k$ up to $K=11$. Note that $\Omega_C$ can be negative, which causes problems for the $K-$th root when considering an even number of hidden units $K$. However a simple permutation of the probability $p$ and $r$ changes the sign of $\Omega_C$. Since the original messages are drawn from the uniform distribution, this permutation can be done without any loss of generality. Instead of using $\mbi{s}_0$, one uses $-\mbi{s}_0$. We did not check higher values of $K$, but we conjecture that the same result holds. This means that the parity tree with non-monotonic hidden units saturates the Shannon bound in the large codeword length limit for any number of hidden units $K$.

\subsection{Tuning of the committee tree with non-monotonic hidden units}
In the committee tree with non monotonic hidden units case, we have
\begin{eqnarray}
\Omega \equiv \Omega_{CTH} = \int_{-\infty}^{+\infty} \left[ \prod_{l=1}^K D z_l \right] \times \sgn \left[ \sum^K_{l=1} f_k (z_l) \right].
\end{eqnarray}
The committee tree with non-monotonic hidden units is optimal if and only if 
\begin{eqnarray}
\Omega_{CTH} = \Omega_{C} \Leftrightarrow \Omega_C & = & \sum_{l=0}^{\frac {K-1} {2}} \binom{K}{l} \left( [2 H(k)]^l [1-2H(k)]^{K-l} \right.
\nonumber \\  && \makebox[3cm]{}
\left. - [2 H(k)]^{K-l} [1-2H(k)]^l \right) \label{kCTH},
\end{eqnarray}
where $\binom{x}{y}$ denotes the binomial coefficient. This gives us a condition on the threshold parameter $k$ of the non-monotonic transfer function $f_k$. If the threshold $k$ is tuned to satisfy (\ref{kCTH}), the scheme achieves the Shannon limit. Thus, we should check if such an optimal threshold $k$ exists. 

We solved (\ref{kCTH}) numerically with parameters $(p,r) \in \{]0,1[\}^2$ and always found an optimal threshold parameter $k$ up to $K=11$. We did not check higher values of $K$, but we conjecture that the same result holds. Note that as mentioned in the definition of this encoder, we considered only an odd number of hidden units $K$. Therefore, these results mean that the committee tree with non-monotonic hidden units saturates the Shannon bound in the large codeword length limit for any odd number of hidden units $K$.

\subsection{Tuning of the parity tree}
In the parity tree case, we have
\begin{eqnarray}
\Omega \equiv \Omega_{PT} = \prod_{l=1}^K \int_{-\infty}^{+\infty} D z_l \times \sgn (z_l).
\end{eqnarray}
The parity tree is optimal if and only if 
\begin{eqnarray}
\Omega_{PT} = \Omega_{C} \Leftrightarrow \Omega_C = 0. \label{kMONOPT}
\end{eqnarray}
This gives us a strong condition on $\Omega_C$. From the definition (\ref{omegC}), it can be easily seen that $\Omega_C=0$ if and only if $r=p$: that is when the BAC channel turns into the particular case of the BSC channel. This means that the standard monotonic parity tree saturates the Shannon bound in the large codeword length limit, but only in the BSC case and for a number of hidden units $K>1$. This confirms what we expected and is the equivalent of Mimura et al. \cite{Mimura2006} lossy compression case.

\FloatBarrier

\section{Conclusion and Discussion} \label{section7}

We investigated an error correcting code scheme for uniformly unbiased Boolean messages using parity tree and committee tree multilayer perceptrons. All the schemes which use the non-monotonic transfer function $f_k$ in their hidden layer were shown to saturate the Shannon bound under some specific conditions. The use of $f_k$ enables the relevant schemes to deal with asymmetric channels like the BAC while monotonic networks using only the standard sign function can deal only with symmetric channels like the BSC.

Indeed, we confirmed that the standard monotonic parity tree saturates the Shannon bound only in the case of the BSC channel. The standard monotonic committee tree however, fails to provide optimal performance even in the BSC case. 

As a general conclusion, this paper shows that tree-like multilayer perceptrons introduced in \cite{Hosaka2002,Mimura2006,Cousseau2008} within the framework of lossy compression can also be used efficiently in an error correcting code scheme. For each network considered, we provided a theoretical analysis of the typical performance and gave the necessary conditions for obtaining optimal performance. In each case, we were able to derive results similar to the lossy compression results. 
Finally, in the case of error correcting code, the replica symmetric solution stability \cite{Almeida1978} was not checked because no replica symmetry breaking is expected on the Nishimori line \cite{Nishimori2001bis}.

This paper discusses only the typical performance of the schemes at the infinite codeword length, however, and does not provide any explicit decoder. 
Because the present schemes make use of densely connected systems, a formal decoder cannot be implemented as it would require a decoding time which would grow exponentially with the size of the original message. One promising alternative is to use the popular belief propagation (BP) algorithm to calculate an approximation of the marginalized posterior probabilities. The BP algorithm is known for giving good results when working in the ferromagnetic phase, where no frustration is present into the system.

With the previous work done on lossy compression \cite{Hosaka2002,Hosaka2006,Mimura2006,Cousseau2008} and on error correcting code \cite{Shinzato2006} using perceptron type networks, there is now a sufficient theoretical background to investigate and compare the practical performance (in the finite codeword length limit) of all the schemes with the theoretical performance. In the case of lossy compression with a simple perceptron, the study of the BP algorithm performance has already been done by Hosaka et al. \cite{Hosaka2006}. Their work provides a solid base from which to begin investigating the more complicated multilayer structure. The influence of the number of hidden units on the practical performance of the scheme is an interesting issue which will be examined in future work.

%%%%%%%%%%%%%%%%%%%%%%%%%%%%%%%%%%%%%%%%%%%%%%%%%%%%%%%%%%%%%%%%%%%%%%%%%%%%%%%%%%%%%%%%%%%%%
%%  ACKNOWLEDGEMENTS

\section*{Acknowledgements}

This work was partially supported by a Grant-in-Aid for Encouragement of
Young Scientists (B) (Grant No. 18700230), Grant-in-Aid for Scientific
Research on Priority Areas (Grant Nos. 18079003, 18020007),
Grant-in-Aid for Scientific Research (C) (Grant No. 16500093), and a
Grant-in-Aid for JSPS Fellows (Grant No. 06J06774) from the Ministry of
Education, Culture, Sports, Science and Technology of Japan.

%%%%%%%%%%%%%%%%%%%%%%%%%%%%%%%%%%%%%%%%%%%%%%%%%%%%%%%%%%%%%%%%%%%%%%%%%%%%%%%%%%%%%%%%%%%%%
%%  APPENDIX
\vspace*{5mm}
\appendix

%%%%%%%%%%%%%%%%%%%%%%%%%%%%%%%%%%%%%%%%%%%%%%%%%%%%%%%%%%%%%%%%%%%%%%%%%%%%%%%%%%%%%%%%%%%%%
%%  APPENDIX A
\section{Analytical Evaluation using the replica method}
\label{appendix.ReplicaMethod}

The free energy can be evaluated by the replica method,
\begin{equation}
f(\beta,R)= - \frac 1 {\beta N} \lim\limits_{n \to 0}
\frac {\left< Z(\beta,\mbi{y},\mbi{x})^n \right>_{\mbi{y},\mbi{x}} - 1} {n}
\end{equation}
where $Z(\beta,\mbi{y},\mbi{x})^n$ denotes the $n$-times replicated partition function
\begin{equation}
Z(\beta,\mbi{y},\mbi{x})^n = \sum_{\mbi{s}^1,\ldots,\mbi{s}^n} \prod_{a =1}^n
\exp \left[ -\beta \mathcal{H} (\mbi{y},\mbi{\hat{y}}(\mbi{s}^{a})) \right]. \label{replicaZ}
\end{equation}
Vector $\mbi{s}^{a}$ is given by
$\mbi{s}^{a}=(\mbi{s}^{a}_1, \ldots ,\mbi{s}^{a}_K)$ and superscript
$a$ denotes the replica index.

We proceed to the calculation of the replicated partition function (\ref{replicaZ}).
Inserting the following two identities,
\begin{eqnarray}
1 & = & \prod_{a=1}^{n} \prod_{l=1}^{K} \int_{-\infty}^{+\infty} d m^{a}_l
\delta \left( \mbi{s}^0_l \cdot \mbi{s}^a_l - \frac N K m^{a}_l \right)
\nonumber \\
& = & \left( \frac 1 {2 \pi i} \right)^{n K}
\int \left( \prod_{a} \prod_{l} d m^{a}_l d \hat{m}^{a}_l  \right)
\nonumber \\
&& \times \exp \left[ \sum_{a} \sum_{l} \hat{m}^{a}_l
 \left( \mbi{s}^0_l \cdot \mbi{s}^a_l - \frac N K m^{a}_l \right) \right] \label{mOrder}
\end{eqnarray}
and
\begin{eqnarray}
1 & = & \prod_{a<b}^{n} \prod_{l=1}^{K} \int_{-\infty}^{+\infty} d q^{ab}_l
\delta \left( \mbi{s}^a_l \cdot \mbi{s}^b_l - \frac N K q^{ab}_l \right)
\nonumber \\
& = & \left( \frac 1 {2 \pi i} \right)^{n(n-1)K/2}
\int \left( \prod_{a<b} \prod_{l} d q^{ab}_l d \hat{q}^{ab}_l  \right)
\nonumber \\
&& \times \exp \left[ \sum_{a<b} \sum_{l} \hat{q}^{ab}_l
 \left( \mbi{s}^a_l \cdot \mbi{s}^b_l - \frac N K q^{ab}_l \right) \right] \label{qOrder}
\end{eqnarray}
into (\ref{replicaZ}) enables us to separate the relevant order parameters, and
to calculate the average moment
$\left< Z(\beta,\mbi{y},\mbi{x})^n \right>_{\mbi{y},\mbi{x}}$ for natural numbers
n as,
\begin{eqnarray}
\left< Z(\beta,\mbi{y},\mbi{x})^n \right>_{\mbi{y},\mbi{x}} & \backsimeq &
\int \left( \prod_{a} \prod_{l} d m^{a}_l \frac {d \hat{m}^{a}_l} {2 \pi i}  \right) \times
\int \left( \prod_{a<b} \prod_{l} d q^{ab}_l \frac {d \hat{q}^{ab}_l} {2 \pi i}  \right)
\nonumber \\
&& \times \exp \left\{ N \left[ R^{-1} \ln \left\{ \sum_{y}
\int \left( \prod_{l} d \mbi{u}_l \frac {d \mbi{v}_l} {2\pi} d R_l \frac {d W_l} {2\pi} \right)
\right. \right. \right. \nonumber \\ &&
\times \left( \frac 1 2 + \frac y 2 \left[ (1-r-p) \mathcal{F}_k ( \{ R_l \} ) + (r-p) \right] \right)
\nonumber \\
&& \times
\prod_{a} \left\{ \exp \left[ \beta \ln \left( \frac 1 2 + \frac y 2 \left[ (1-r-p) \mathcal{F}_k ( \{ u^a_l \} ) + (r-p) \right] \right) \right] \right\}
\nonumber \\
&&
\times \prod_{l} \Bigg\{ \exp \Bigg[ - \frac 1 2 (W_l)^2 - \frac 1 2 \mbi{v}_l \cdot \mathcal{Q}_l \cdot \mbi{v}_l - W_l \mathcal{M}_l \cdot \mbi{v}_l 
\nonumber \\ &&
\makebox[2cm]{} + i R_l W_l + i \mbi{v}_l \cdot \mbi{u}_l \Bigg] \Bigg\} \Bigg\}
\nonumber \\ &&
+ \frac 1 K \ln \left\{ \sum_{s^a} \exp \left[ \sum_{a,l} \hat{m}^a_l s^a_l + \sum_{a<b,l} \hat{q}^{ab}_l s^a_l s^b_l \right] \right\}
\nonumber \\ &&
- \frac 1 K \sum_{a,l} m^a_l \hat{m}^a_l  - \frac 1 K \sum_{a<b,l} q^{ab}_l \hat{q}^{ab}_l
- \beta \sum_a \ln 2 \Bigg] \Bigg\}, \label{zgeneral}
\end{eqnarray}
where $\mathcal{Q}_l$ is an $n \times n$ matrix having elements $\{ q^{ab}_l \}$
and where $\mathcal{M}_l$ is an $n$ dimensional vector having elements $\{ m^a_l \}$.
We analyze the scheme at the thermodynamic limit $N,M \to + \infty$ while the code rate $R$ is kept finite. In this limit, (\ref{zgeneral}) can be evaluated using the saddle point method with respect to $m^a,\hat{m}^a,q^{ab}_l,\hat{q}^{ab}_l$ so that the free energy can be retrieved. To continue the calculation, we have to make some assumptions about the structure of these order parameters. In this paper, we use the so-called replica symmetric (RS) ansatz,
\begin{equation}
\begin{array}{ccc}
m^a_l = m, & \quad & q^{ab}_l = (1-q) \delta_{ab} + q,
\\
\hat{m}^a_l = \hat{m}, & \quad & \hat{q}^{ab}_l = (1- \hat{q} ) \delta_{ab} + \hat{q},
\end{array}
\label{RS}
\end{equation}
where $\delta_{ab}$ denotes the Kronecker delta. This ansatz means that all the hidden
units are equivalent after averaging over the disorder.

Also note that by definition, order parameter $m$ is equivalent to quantity $\frac{\mbi{s}^0 \cdot \mbi{s}} N$, which gives the overlap between the decoded message $\mbi{s}$ and the original message $\mbi{s}^0$. An overlap of $1$ indicates perfect decoding while an overlap of $0$ denotes complete failure.

%%%%%%%%%%%%%%%%%%%%%%%%%%%%%%%%%%%%%%%%%%%%%%%%%%%%%%%%%%%%%%%%%%%%%%%%%%%%%%%%%%%%%%%%%%%%%
%%%%%%%%%%%%%%%%%%%%%%%%%%%%%%%%%%%%%%%%%%%%%%%%%%%%%%%%%%%%%%%%%%%%%%%%%%%%%%%%%%%%%%%%%%%%%

%%%%%%%%%%%%%%%%%%%%%%%%%%%%%%%%%%%%%%%%%%%%%%%%%%%%%%%%%%%%%%%%%%%%%%%%%%%%%%%%%%%%%%%%%%%%%
\end{document}